\documentclass[ reprint,nofootinbib,amsmath,amssymb,aps]{revtex4-2}
\usepackage{appendix}
\usepackage{graphicx}
\usepackage{dcolumn}
\usepackage[colorlinks,linkcolor=magenta,anchorcolor=cyan,citecolor=blue]{hyperref}
\usepackage{physics}
\usepackage{bm}
\usepackage[multiple]{footmisc}

\def\be{\begin{equation}}
\def\ee{\end{equation}}
\def\ba{\begin{eqnarray}}
\def\ea{\end{eqnarray}}

\def\dbar{{\mathchar '26\mkern -10mu\delta}}

              \def\.{\cdot}

\begin{document}
\title{ Equivalence of Noether charge and Hilbert action boundary term formulas for the black hole entropy in $F(R_{abcd})$ gravity theory}
\author{Wei Guo$^{1,2}$}
\email{guow@mail.bnu.edu.cn}
\author{Xiyao Guo$^{1,2}$}
\email{xiyaoguo@mail.bnu.edu.cn}
\author{Mingfeng Li$^{1,2}$}
\email{mingfengli@mail.bnu.edu.cn}
\author{Zili Mou$^{1,2}$}
\email{zilimou@mail.bnu.edu.cn}
\author{Hongbao Zhang$^{1,2}$}
\email{hongbaozhang@bnu.edu.cn}
\affiliation{$^1$School of Physics and Astronomy, Beijing Normal University, Beijing 100875, China\label{addr1}\\
$^2$ Key Laboratory of Multiscale Spin Physics, Ministry of Education, Beijing Normal University, Beijing 100875, China}

\date{\today}

\begin{abstract}
By working with the covariant phase space formalism, we have shown that not only can the Hamiltonian conjugate to a Killing vector field $\xi$ be expressed as the sum of the associated Noether charge and $\xi$ contracted with the Hilbert action boundary term for $F(R_{abcd})$ gravity, but also be written as its contraction with another $\xi$ independent tensor field. With this, we have proven the equivalence of Noether charge and Hilbert action boundary term formulae for the stationary black hole entropy in $F(R_{abcd})$ gravity, which is further substantiated by our explicit computation using both formulae. 
\end{abstract}
\maketitle
\section{Introduction}
Long before, the first law of black hole mechanics was derived for an arbitrary diffeomorphism invariant Lagrangian theory of gravity in \cite{Wald,IW}, whereby the black hole entropy can be calculated as a Noether charge by Wald formula, which 
reduces to the famous Bekeinstein-Hawking formula for Einstein's general relativity. Later on, the comparison of the aforementioned Noether charge approach with various Euclidean methods was further made in \cite{Iyer}, where it was shown that all of these Euclidean methods within their domains of applicability give rise to exactly the same result as that by Noether charge approach. In particular, the authors in \cite{Iyer} widened the domain of applicability of BTZ prescription proposed in \cite{BTZ} by making two ansatzes. One is assuming that the Hamiltonian $H_\xi$ conjugate to an arbitrary vector field $\xi$ tangential to the boundary can be expressed as the sum of Noether charge and $\xi$ contracted with Hilbert action boundary term. The other is assuming that the above Hamiltonian can also be expressed in terms of $\xi$ contracted with another $\xi$ independent tensor field. These two ansatzes were further explicitly verified to be satisfied by Einstein's general relativity. However, such an assumption is surely a sufficient but not an a priori necessary condition for the proof of the equivalence of Noether charge and Hilbert action boundary term formulae for the stationary black hole entropy. Actually to fulfill the above proof,  the aforementioned assumption is required to be valid only for the time translation Killing vector field normal to the event horizon. The purpose of this paper is to investigate such an equivalence for $F(R_{abcd})$ gravity. As a result, we find that both ansatzes are not satisfied by generic $F(R_{abcd})$ gravity but satisfied when restricted onto the Killing vector fields, including the time translation Killing vector field normal to the event horizon. Thus the above equivalence can be formally proven. Moreover, we also demonstrate such an equivalence by using both approaches to explicitly calculate the stationary black hole entropy. 

The structure of this paper is organized as follows. In the subsequent section, we shall work with the covariant phase space formalism not only to derive the explicit expression for a variety of relevant quantities such as Noether charge and Hilbert action boundary term, but also to show why the aforementioned assumption is satisfied at least by the Killing vector fields for $F(R_{abcd})$ gravity. Then in Section \ref{NH}, we will prove the equivalence of Noether charge and Hilbert action boundary term formulae for the stationary black hole entropy in $F(R_{abcd})$ gravity by both formal derivation and explicit calculation. We conclude our paper with some discussions. 

We will follow the notation and conventions of \cite{GR}. In particular, we shall use the boldface letters to denote differential forms with the tensor indices suppressed. 
\section{Covariant phase space formalism}
Let us start from the following Lagrangian form for $F(R_{abcd})$ gravity\cite{DSSY,JZ}, i.e.,
\begin{equation}
\mathbf{L}=\bm{\epsilon}\left[F(\varrho_{abcd},g_{ab})+\psi^{abcd}(R_{abcd}-\varrho_{abcd})\right],
\end{equation}
where $\bm{\epsilon}$ is the spacetime volume form, and $F$ is an arbitrary function of $\varrho_{abcd}$ and $g_{ab}$ with $\varrho_{abcd}$ and $\psi^{abcd}$ two auxiliary fields, both of which have all the symmetries of Riemann tensor $R_{abcd}$. Whence we have
\begin{equation}
    \delta \mathbf{L}= \bm\epsilon E_\phi\delta\phi+d\mathbf{\Theta}(\phi,\delta\phi).
\end{equation}
Here the dynamical variables are denoted collectively by $\phi$ with 
\begin{eqnarray}
E_g^{ab}&=&\frac{1}{2}g^{ab}F+\frac{1}{2}g^{ab}\psi^{cdef}(R_{cdef}-\varrho_{cdef})\nonumber\\ & &+\frac{\partial F}{\partial g_{ab}}+2\nabla_{c}\nabla_{d}\psi^{c(ab)d}+\psi^{cde(a}R_{cde}{}^{b)},\\
E_{\psi abcd}&=& R_{abcd}-\varrho_{abcd},\\
E_\varrho^{abcd}&=& P^{abcd}-\psi^{abcd},\\
\mathbf{\Theta}&=&\theta\cdot\bm{\epsilon},
\end{eqnarray}
where 
$ P^{abcd}\equiv\pdv{F}{\varrho_{abcd}}$,
and \begin{equation}
\theta^a=2(\nabla_d\psi^{bdca}\delta g_{bc}-\psi^{bdca}\nabla_d\delta g_{bc})
\end{equation}
with the dot representing the contraction of a vector index and the first index of a differential form. It is noteworthy that all quantities presented in this section are exactly the same as those derived from the original Lagrangian form $\mathbf{L}=F(R_{abcd},g_{ab})\bm\epsilon$ for $F(R_{abcd})$ gravity when $E_{\psi abcd}=E_\varrho^{abcd}=0$.

The Noether current associated with the diffeomorphisms generated by a vector field $\xi$ is defined as follows
\begin{equation}\label{Noethercurrent}
\mathbf{J}_\xi=X_\xi\cdot \mathbf{\Theta}-\xi\cdot \mathbf{L},
\end{equation}
where $X_\xi=\int d^dx\sqrt{-g}\mathcal{L}_\xi\phi(x)\frac{\delta }{\delta\phi(x)}$ is understood as a vector at the point $\phi(x)$ in the configuration space and $\delta$ is thought of as the exterior derivative in the configuration space. Such a Noether current is conserved as it should be the case when the equations of motion are satisfied because 
\begin{eqnarray}\label{nice}
d\mathbf{J}_\xi=-\bm\epsilon E_\phi\mathcal{L}_\xi\phi.
\end{eqnarray}
It then follows from Eq. (\ref{nice}) by an explicit calculation that 
\begin{equation}
    d(\mathbf{J}_\xi-\bm{\mathcal{C}}_\xi)=\bm\epsilon \xi^f\mathcal{B}_f,
\end{equation}
where 
\begin{equation}
    \bm{\mathcal{C}}_\xi=c_\xi\cdot \bm\epsilon
\end{equation}
with 
\begin{equation}
    c_\xi^a=\xi^f(-2E_g^a{}_f+4\psi^{abcd}E_{\psi fbcd}-4E_\varrho^{abcd}\varrho_{fbcd})
\end{equation}
and 
\begin{eqnarray}
    \mathcal{B}_f&=&2\nabla_aE_g^a{}_f-E_{\psi abcd}\nabla_f\psi^{abcd}-4\nabla_a(E_{\psi fbcd}\psi^{abcd})\nonumber\\
   && -E_\varrho^{abcd}\nabla_f\varrho_{abcd}+4\nabla_a(E_\varrho^{abcd}\varrho_{fbcd}).
\end{eqnarray}
The Bianchi identity $\mathcal{B}_f=0$ can be obtained by taking into account the fact that $\xi$ can be chosen in an arbitrary manner. Accordingly, the Noether current can be expressed as 
\begin{eqnarray}
\mathbf{J}_\xi=d\mathbf{Q}_\xi+\bm{\mathcal{C}}_\xi,
\end{eqnarray}
where, as calculated explicitly in  Appendix \ref{Noethercharge}, the Noether charge can be  constructed locally as 
\begin{equation}
 \mathbf{Q}_\xi=(-\psi^{cadb}\nabla_{[d}\xi_{b]}+2\nabla_{[d}\psi^{cadb}\xi_{b]})\bm{\epsilon}_{ca\cdot\cdot\cdot}
\end{equation}

Now let us consider the spacetime $M$ with a boundary $\partial M$, which can be either spacelike or timelike. The induced metric on the boundary can be written as
\begin{equation}
    h_{ab}=g_{ab}-\varepsilon n_an_b
\end{equation}
with $n^a$ chosen to be the outer-directed normal vector if it is spacelike and the past-directed normal vector if it is timelike.  Suppose that the boundary is fixed under variation, then we have 
\begin{equation} 
\delta n_a=\delta a n_a,
\end{equation}
where $\delta a=-\frac{\varepsilon}{2}\delta g^{ab}n_an_b$ can be obtained by $g^{ab}n_an_b=\varepsilon=\pm1$ and $\delta(g^{ab}n_an_b)=0$. Whence it is not hard to further show
\begin{eqnarray}
    \delta n^a=\delta g^{ab}n_b+g^{ab}\delta n_b=-\delta a n^a-\varepsilon \dbar A^a,
\end{eqnarray}
where $\dbar A^a=-\varepsilon h^a{}_b\delta g^{bc} n_c$.
Accordingly, the variation of the metric on the boundary can be expressed as 
\begin{equation}
    \delta g^{ab}=-2\varepsilon \delta a n^an^b-\dbar A^a n^b-n^a\dbar A^b+\delta h^{ab}.
\end{equation}
Then by a tedious but straightforward calculation, one winds up with\cite{JZ}
\begin{eqnarray}\label{corner}
    &&\mathbf{\Theta}|_{\partial M}=\hat{\bm\epsilon}[-4\varepsilon \Psi_{ab}\delta K^{ab}-2D^a(\Psi_{ab}\dbar A^b)+\nonumber\\
   && (2n^a\nabla^e\psi_{beac}+6\varepsilon\Psi_{ab}K^a{}_c)\delta h^{bc}-2n^d\psi_{cadb}D^a\delta h^{bc}],\nonumber\\
\end{eqnarray}
where $K_{ab}$ represents the extrinsic curvature, $\Psi_{ab}=\psi_{acbd}n^cn^d$, $D_a$ is the boundary covariant derivative, and $\hat{\bm\epsilon}$ is the induced volume element on the boundary, defined as $\bm\epsilon=\mathbf n\wedge \hat{\bm\epsilon}$ as usual. The above equation can be further cast into the following form, i.e.,
\begin{eqnarray}
    \mathbf{\Theta}|_{\partial M}=-\delta \mathbf{B}+d\mathbf{C}+\mathbf{F},
\end{eqnarray}
where
\begin{eqnarray}
    \mathbf{B}&=&4\varepsilon\Psi_{ab}K^{ab}\hat{\bm\epsilon},\nonumber\\
    \mathbf{C}&=&\mathbf\omega\cdot\hat{\bm\epsilon},\nonumber\\
    \mathbf{F}&=&\hat{\bm\epsilon}(T_{hbc}\delta h^{bc}+T_{\Psi bc}\delta\Psi^{bc})
\end{eqnarray}
with
\begin{eqnarray}\label{bc}
    \omega^a&=&-2\Psi^a{}_b\dbar A^b+2h^{ae}\psi_{ecdb}n^d\delta h^{bc},\nonumber\\
    T_{hbc}&=&-2\varepsilon\Psi_{de}K^{de}h_{bc}+2n^a\nabla^e\psi_{deaf}h^d{}_{(b}h^f{}_{c)}\nonumber\\
    &&-2\varepsilon \Psi_{a(b}K^a{}_{c)}-2D^a(h_a{} ^eh_{(c}{}^f\psi_{|efd|b)}n^d),\nonumber\\
    T_{\Psi bc}&=&4\varepsilon K_{bc}.
\end{eqnarray}
Next let us focus on the case in which the $\xi$ is tangential to $\partial M$, whereby we have 
\begin{eqnarray}\label{new1}
    \int_{\partial M}( \mathbf{J}_\xi+X_\xi\cdot \delta \mathbf{B})&=&\int_{\partial M}[-\xi\cdot \mathbf{L}+d(X_\xi\cdot \mathbf{C})+X_\xi\cdot \mathbf{F}]\nonumber\\
   &=& \int_{\partial M}[d(X_\xi\cdot \mathbf{C})+X_\xi\cdot \mathbf{F}].
\end{eqnarray}
On the other hand, we also have
\begin{eqnarray}\label{new2}
     \int_{\partial M}( \mathbf{J}_\xi+X_\xi\cdot \delta \mathbf{B})&=&\int_{\partial M}(\bm{\mathcal{C}}_\xi+d\mathbf{Q}_\xi+\mathcal{L}_\xi \mathbf{B})\nonumber\\
     &=&\int_{\partial M}[\bm{\mathcal{C}}_\xi+d(\mathbf{Q}_\xi+\xi\cdot \mathbf{B})],
\end{eqnarray}
where we have used the diffeomorphism covariance of $\mathbf{B}$ in the sense that $\mathcal{L}_{X_\xi}\mathbf{B}=\mathcal{L}_\xi\mathbf{B}$.
Combining Eq. (\ref{new1}) and Eq. (\ref{new2}), we further have
\begin{eqnarray}
  && \int_{\partial M}d(\mathbf{Q}_\xi+\xi\cdot \mathbf{B}-X_\xi\cdot \mathbf{C})=\int_{\partial M}(X_\xi\cdot \mathbf{F}-\bm{\mathcal{C}}_\xi)\nonumber\\
   &&=\int_{\partial M}[d(q_\xi \cdot \hat{\bm\epsilon})+\xi^f\Lambda_f\hat{\bm\epsilon}-\bm{\mathcal{C}}_\xi],
\end{eqnarray}
where 
\begin{eqnarray}\label{brownyork}
    q_\xi^a&=&-(2T_h{}^a{}_c+2\Psi^{ab}T_{\Psi cb})\xi^c,\nonumber\\
    \Lambda_f&=&2D^bT_{hbf}+T_{\Psi bc}D_f\Psi^{bc}
+2D_d(T_{\Psi fc}\Psi^{dc})
\end{eqnarray}
is obtained by the fact that $X_\xi\cdot \delta h^{bc}=-2D^{(b}\xi^{c)}$ and $X_\xi\cdot\Psi^{bc}=\xi^dD_d\Psi^{bc}-\Psi^{dc}D_d\xi^b-\Psi^{bd}D_d\xi^c$ as well as the integration by parts. 
The arbitrariness of $ \xi$ implies the following two identities, i.e., 
\begin{equation}
    \Lambda_f=n_a(-2E_g^a{}_f+4\psi^{abcd}E_{\psi fbcd}-4E_\varrho^{abcd}\varrho_{fbcd}),
\end{equation}
and 
\begin{equation}\label{imp1}
    \int _\mathcal{S} (\mathbf{Q}_\xi+\xi\cdot \mathbf{B}-X_\xi\cdot \mathbf{C})=\int_\mathcal{S} q_\xi\cdot\hat{\bm\epsilon},
\end{equation}
where $\mathcal{S}$ is any closed co-dimension 2 cross section on $\partial M$.
\begin{figure}
\centering
\begin{minipage}{0.02\textwidth}
  \ \ \\
  \end{minipage}
\includegraphics[width=0.5\textwidth]{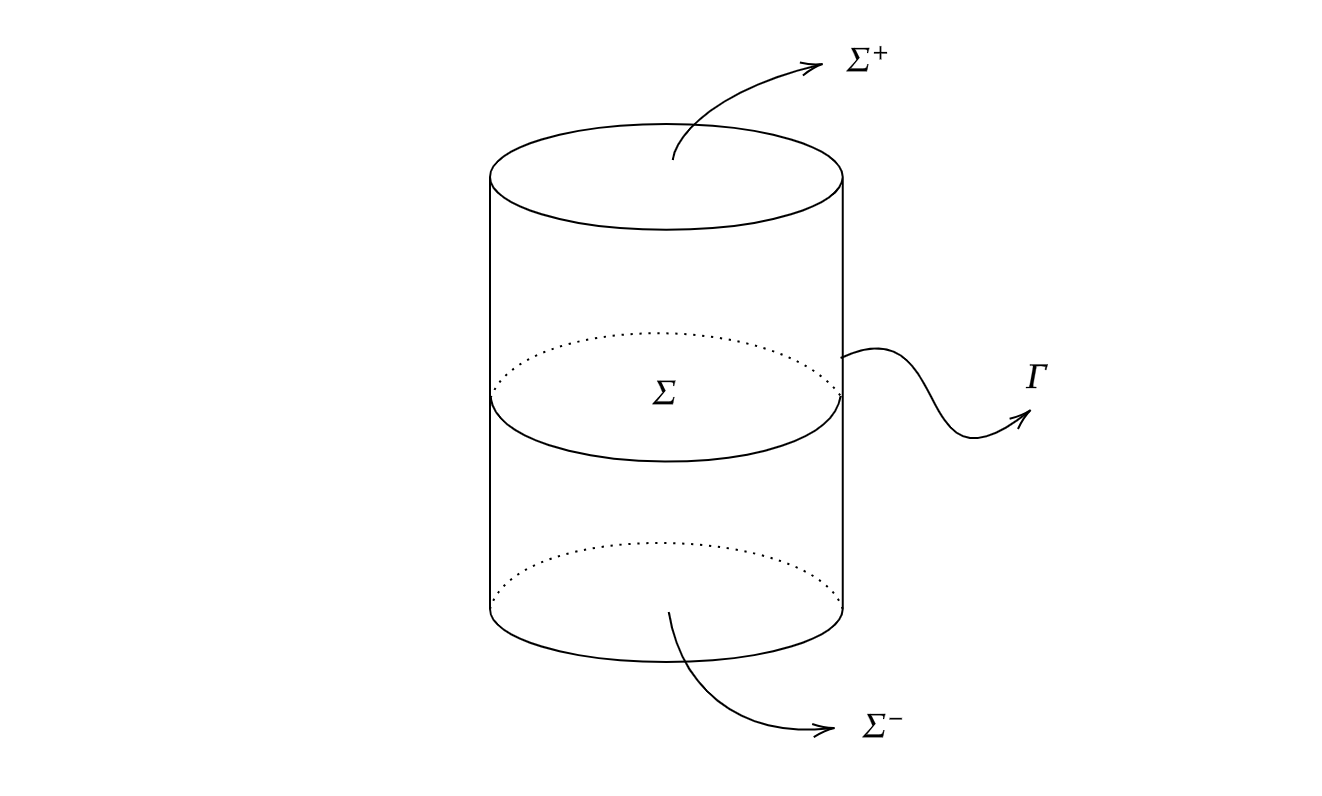}
\caption{The spacetime region $M$ is bounded by a timelike surface $\Gamma$ and sandwiched by two spacelike surfaces $\Sigma_\pm$ with another spacelike surface $\Sigma$ in between.}\label{fig1}
\end{figure}

As shown in Figure \ref{fig1}, we further restrict ourselves onto the spacetime region $M$ sandwiched by two spacelike surfaces $\Sigma_\pm$ and bounded by a timelike surface $\Gamma$. Although $\Sigma_\pm$ and $\Gamma$ serve as the portion of the boundary of the spacetime we are considering, they play totally distinct roles, where the data on $\Sigma_\pm$ encode the information about the initial value of the theory, while $\Gamma$ is the true boundary for one to impose some appropriate boundary condition over there. In particular, with the boundary condition $\delta h^{ab}=\delta \Psi^{ab}=0$ on $\Gamma$, the $\mathbf{F}$ term vanishes and only the first term in Eq. (\ref{bc}) survives for $\mathbf{C}$, which we denote as $\mathbf{C}_\Gamma$. As a result, the Hilbert action, supplemented with the boundary term $\mathbf{B}$ as $I=\int_M\mathbf{L}+\int_{\partial M}\mathbf{B}$, will be extremized by solutions to the equations of motion. On the other hand, together with the fact that  $d\delta \mathbf{\Theta}=\delta^2 \mathbf{L}-\bm\epsilon \delta E_\phi \wedge \delta \phi=0$ when restricted onto the solution space, one can show that the corresponding symplectic form, defined as follows\cite{Harlow,Zhang}
\begin{equation}
    \Omega =\int_{\Sigma}\delta \mathbf{\Theta}-\int_{\partial \Sigma}\delta \mathbf{C}_\Gamma,
\end{equation}
does not depend on the choice of the spacelike surface $\Sigma$. Furthermore, for $\xi$ tangential to $\Gamma$, we  can arrive at 
\begin{equation}
    -X_\xi \cdot \Omega=\delta \int_\Sigma \mathbf{J}_\xi+ \delta \int_{\partial_\Sigma}(\xi\cdot\mathbf{B}-X_\xi\cdot \mathbf{C}_\Gamma)
\end{equation}
by the diffeomorphism covariance of $\mathbf{\Theta}$ and $\mathbf{C}_\Gamma$. 
 This motivates one to define the Hamiltonian conjugate to $\xi$ as
 \begin{equation}
     H_\xi(\Sigma)=\int_\Sigma \mathbf{J}_\xi+ \int_{\partial \Sigma}(\xi\cdot\mathbf{B}-X_\xi\cdot\mathbf{C}_\Gamma),
 \end{equation}
 which reduces to a pure surface term as follows 
 \begin{equation}\label{imp2}
     H_\xi(\Sigma)=\int_{\partial \Sigma}(\mathbf{Q}_\xi+\xi\cdot\mathbf{B}-X_\xi\cdot\mathbf{C}_\Gamma)
 \end{equation}
 when evaluated on the solution space. 
 
 Note that $\psi_{abcd}=\frac{1}{2}(g_{ac}g_{bd}-g_{ad}g_{bc})$ and $\Psi_{ab}=\frac{1}{2}h_{ab}$ for Einstein's general relativity. As a result, the second term in Eq. (\ref{bc}) for $\mathbf{C}$ vanishes automatically with $\omega^a=-\dbar A^a$.  If $\Sigma$ is chosen to be orthogonal to $\Gamma$ and is required to keep such an orthogonality under variation, then both $n^a$ and $\dbar A^a$ are tangential to $\Sigma$, which implies $\mathbf{C}_\Gamma=\mathbf{C}=0$ when restricted onto $\partial \Sigma$.  Accordingly, it follows from Eq. (\ref{imp1}) and Eq. (\ref{imp2}) that the resulting Hamiltonian can be written as\cite{Iyer}
\begin{equation}\label{fundamental}
    H_\xi(\Sigma)=\int_{\partial\Sigma}(\mathbf{Q}_\xi+\xi\cdot \mathbf{B})=\int_{\partial \Sigma}q_\xi\cdot\hat{\bm\epsilon}.
\end{equation}
Apparently, the above result is not guaranteed to be valid for generic $F(R_{abcd})$ gravity with $\xi$ a generic vector field tangential to $\Gamma$. But nevertheless, if $\xi$ is a Killing vector field, then we have $X_\xi\cdot \mathbf{C}_\Gamma=X_\xi\cdot \mathbf{C}=0$ at $\partial\Sigma$ even without requiring $\Sigma$ to be orthogonal to $\Gamma$. Thus we can still have Eq. (\ref{fundamental}), where $q_\xi\cdot\hat{\bm{\epsilon}}$ can be viewed as the generalization to $F(R_{abcd})$ gravity of the Brown-York local charge density associated with $\xi$\cite{BY}.

\section{The equivalence of Noether charge and Hilbert action boundary term formulae}\label{NH}
In this section, we shall argue for the equivalence of Noether charge and Hilbert action boundary term formulae for the stationary black hole entropy, namely\footnote{Our expression for the BTZ prescription is different from that presented in \cite{Iyer} because  we are using different prescriptions in performing the analytic continuation to the Euclidean sector.}
\begin{equation}\label{equivalence}
   \frac{2\pi}{\kappa}\int_\mathcal{B}\mathbf{Q}_{\frac{\partial}{\partial t}}=\lim_{r\rightarrow 0}\int_{\partial D_r\times \mathcal{B}} -i\mathbf{B}.
\end{equation}
The left handed side of this equality is the Wald formula\cite{Wald,IW}, where $\frac{\partial}{\partial t}$ is the time translation Killing vector field with the surface gravity $\kappa$ on the event horizon and $\mathcal{B}$ is the bifurcation surface. The right handed side is the BTZ prescription\cite{BTZ}, where $D_r$ is a two dimensional disk of radius $r$ transverse to the bifurcation surface in the corresponding Euclidean black hole with $t$ replaced by $-i\tau$, and the boundary of $D_r$ is given by the orbit of $\frac{\partial}{\partial \tau}$.

First, as explicitly written later on in Eq. (\ref{why}), $\frac{\partial}{\partial t}$ vanishes at the bifurcation surface. Second, as demonstrated in the first equation of Eq. (\ref{brownyork}), $q_\xi$ is linear in $\xi$ without any derivative acting on it. So it follows from Eq. (\ref{fundamental}) that
\begin{equation}
  \lim_{r\rightarrow 0}  \int_{\partial\Sigma_{0r} }\mathbf{Q}_{\frac{\partial}{\partial t}}=-\lim_{r\rightarrow 0}\int_{\partial\Sigma_{0r}}\frac{\partial}{\partial t}\cdot \mathbf{B},
\end{equation}
where $\Sigma_{0r}$ is the constant $t=0$ surface extended from the bifurcation surface all the way to the boundary of radius $r$. Thus we further have 
\begin{eqnarray}
    \frac{2\pi}{\kappa}\int_\mathcal{B}\mathbf{Q}_{\frac{\partial}{\partial t}}&=&-\lim_{r\rightarrow 0}\frac{2\pi}{\kappa}\int_{\partial\Sigma_{0r}}\frac{\partial}{\partial t}\cdot\mathbf{B}\nonumber\\
    &=&\lim_{r\rightarrow 0}\frac{2\pi i}{\kappa}\int_{\partial \Sigma_{0r}}\frac{\partial }{\partial\tau}\cdot\mathbf{B}\nonumber\\
    &=&-i\lim_{r\rightarrow 0}\int_{\partial D_r\times \mathcal{B}}d\tau\wedge \frac{\partial}{\partial \tau}\cdot \mathbf{B}\nonumber\\
    &=&\lim_{r\rightarrow 0}\int_{\partial D_r\times\mathcal{B}}-i \mathbf{B},
\end{eqnarray}
where the analytic continuation to the Euclidean sector has been made in the second step and the periodicity along $\tau$ to avoid the conical singularity has been used in the third step, with the minus sign removed due to our convention for the orientation.

Next we shall perform the explicit computation for both sides of Eq. (\ref{equivalence}) to show they are exactly equal to each other as it should be the case. To this end, we like to work with the Gaussian null coordinates in the neighbourhood of the event horizon, where the metric reads
\begin{equation}
    ds^2=2dudv-u^2\alpha dv^2+2uw_idvdx^i+\gamma_{ij}dx^idx^j,
\end{equation}
and the inverse metric reads
\begin{equation}
g^{\mu\nu}\partial_\mu\partial_\nu=u^2(\alpha+w^2)\partial_u^2+2\partial_u\partial_v-2uw^i\partial_u\partial_i+\gamma^{ij}\partial_i\partial_j
\end{equation}
with $w^i=\gamma^{ij}w_j$ and $w^2=w_iw_j\gamma^{ij}$. Here $u$ is chosen to be the affine parameter in the whole neighbourhood with $u=0$ on the event horizon, while $v$ is chosen to be the affine parameter only on the event horizon. Furthermore, we require 
$\alpha$, $w_i$, and $\gamma_{ij}$ be functions of $r$ and $x^i$ with $r=uv$ and the bifurcation surface be located at $u=v=0$. This amounts to saying that the time translation Killing vector field takes the following form
\begin{equation}\label{why}
   \xi= \frac{\partial}{\partial t}=\kappa(v\frac{\partial}{\partial v}-u\frac{\partial}{\partial u})
\end{equation}
with $t=\frac{1}{2\kappa}\ln\frac{v}{u}$. Instead of working directly with $\partial D_r\times \mathcal{B}$, we first work with its Lorentz version $\Gamma _r$, which is given by the surface of constant $r$. The resulting induced metric reads
\begin{equation}
    d\hat{s}^2=-\kappa^2r(r\alpha+2)dt^2+2\kappa rw_idtdx^i+\gamma_{ij}dx^idx^j,
\end{equation}
whereby the determinant of the induced metric can be obtained as\footnote{Here we have used the matrix determinant identity $\smdet{
    A & B\\
    C& D}=\smdet{D}\smdet{A-BD^{-1}C}$, which can be obtained by the fact that $\smqty(
    1&-BD^{-1}\\
    0&1
)\smqty(
    A&B\\
    C&D
)=\smqty(
    A-BD^{-1}C& 0\\
    C&D
)$.}
\begin{equation}
   \sqrt{-h}=\frac{\kappa}{\lambda}\sqrt{\gamma}
\end{equation}
with
\begin{equation}
    \lambda=\sqrt{\frac{1}{r[r(\alpha+w^2)+2]}}.
\end{equation}
Accordingly, we have 
\begin{equation}
    \hat{\bm\epsilon}=-\frac{\kappa}{\lambda}dt\wedge\tilde{\bm\epsilon}
\end{equation}
with $\tilde{\bm\epsilon}$ the volume on the $\partial\Sigma_{tr}$, where $\Sigma_{tr}$ is defined in a similar manner as $\Sigma_{0r}$.
In addition, the normal vector to $\Gamma_r$ can be calculated out as
\begin{eqnarray}
    n_a&=&\lambda (dr)_a=\lambda[u(dv)_a+v(du)_a]\nonumber\\
    &\xrightarrow[r\rightarrow 0]{t=0}&\sqrt{\frac{1}{2}}[(du)_a+(dv)_a]+\mathcal{O}(r).
\end{eqnarray}
Then according to the leading behavior of $\nabla_an_b$ given in Appendix \ref{leading} as $r\rightarrow 0$ along the $t=0$ surface, we have
\begin{eqnarray}
\mathbf{B}&=&4\psi^{acbd}n_cn_dK_{ab}\hat{\bm\epsilon}=-\frac{4\kappa}{\lambda}\psi^{acbd}n_cn_d\nabla_{a}n_{b}dt\wedge \tilde{\bm\epsilon}\nonumber\\
&\underset{r\rightarrow 0}\longrightarrow&4\kappa \psi^{uvuv} dt\wedge\tilde{\bm\epsilon}
\end{eqnarray}
at $\partial \Sigma_{0r}$.
Now with the analytic continuation to the Euclidean sector, we have $-i\mathbf{B}=4\kappa \psi^{uvuv}d\tau\wedge\tilde{\bm\epsilon}$, then the black hole entropy by BTZ prescription, namely via the right handed side of Eq. (\ref{equivalence}), can be expressed explicitly as
\begin{equation}\label{BTZfinal}
    S=-8\pi\int_\mathcal{B}\psi^{uvuv}\tilde{\bm\epsilon},
\end{equation}
where we have used the fact that the orientation on $\partial D_r\times \mathcal{B}$ is given by $-d\tau\wedge \tilde{\bm\epsilon}$.
On the other hand, note that $\xi=0$ and
\begin{equation}
    d\bm{\xi}=2\kappa dv\wedge du,\quad \epsilon=dv\wedge du\wedge \tilde{\bm\epsilon}
\end{equation}
at the bifurcation surface, so the black hole entropy by Wald formula, namely via the left handed side of Eq. (\ref{equivalence}), can be calculated explicitly as 
\begin{equation}\label{Waldfinal}
    S=\frac{2\pi}{\kappa}\int_\mathcal{B}(-4\kappa \psi^{vuvu})\tilde{\bm\epsilon}=-8\pi\int_\mathcal{B}\psi^{uvuv}\tilde{\bm\epsilon}.
\end{equation}
Eq. (\ref{BTZfinal}) and Eq. (\ref{Waldfinal}) tells us that the Noether charge and Hilbert action boundary term formulae give rise to exactly the same result for the stationary black hole entropy in $F(R_{abcd})$ gravity.  So we are done. 

\section{Conclusion}
For a Killing vector field $\xi$ tangential to the boundary, we have shown that not only can the corresponding Hamiltonian be written as the sum of the associated Noether charge and its contraction with the Hilbert action boundary term for $F(R_{abcd})$ gravity, but also can be expressed equally as $\xi$ contracted with another $\xi$ independent tensor field. Whence we have succeeded in formally deriving the equivalence of Noether charge and Hilbert action boundary term formulae for the stationary black hole entropy in $F(R_{abcd})$ gravity, which is further confirmed by our explicit calculation using both formulae. 

However, an arbitrary diffeomorphism invariant Lagrangian theory of gravity also involves the derivatives of Riemann tensor. Then a natural question is whether the above equivalence also applies to such a more general case. For the moment, we have no answer to it, but we expect to apply the whole machinery developed here to address this issue elsewhere in the near future.
\begin{acknowledgments}
This work is partly supported by the National Key Research and Development Program of China with Grant No. 2021YFC2203001 as well as the National Natural Science Foundation of China with Grant Nos. 12075026 and 12361141825. H.Z. also likes to acknowledge Zhengwen Liu for his wonderful hospitality at Shing-Tung Yau Center of Southeast University during the final stage of this work.

\end{acknowledgments}

\newpage

\onecolumngrid
\appendix
\section{An explicit calculation of the Noether charge for $F(R_{abcd})$ gravity}\label{Noethercharge}
Here we would like to present an explicit calculation of the Noether charge for $F(R_{abcd})$ gravity. As such, we have 
$\mathbf{J}_\xi-\bm{\mathcal{C}}_\xi=\mathbf\mu\cdot \bm\epsilon$ with 
\begin{eqnarray}
\mu^a&=&2\nabla_d\psi^{bdca}(\nabla_b\xi_c+\nabla_c\xi_b)-2\psi^{bdca}\nabla_d(\nabla_b\xi_c+\nabla_c\xi_b)+2\xi_b\frac{\partial F}{\partial g_{ab}}+4\xi_b\nabla_c\nabla_d\psi^{c(ab)d}+2\psi^{cde(a}R_{cde}{}^{b)}\xi_b\nonumber\\
&&-4\xi^f\psi^{abcd}R_{fbcd}+4\xi^fP^{abcd}\varrho_{fbcd}\nonumber\\
&=&2\nabla_d\psi^{bdca}(\nabla_b\xi_c+\nabla_c\xi_b)+\psi^{dbca}R_{dbcf}\xi^f+2\psi^{dcba}\nabla_d\nabla_c\xi_b+2\psi^{cbda}\nabla_d\nabla_c\xi_b-2\nabla_c(\nabla_d\psi^{cadb}\xi_b)\nonumber\\
&&+2\nabla_d\psi^{cadb}\nabla_c\xi_b-2\xi_b\nabla_c\nabla_d\psi^{bacd}-2\xi_b\nabla_c\nabla_d\psi^{acbd}+\psi^{cdeb}R_{cde}{}^a\xi_b-3\xi^f\psi^{abcd}R_{fbcd}+2\nu_\xi^a\nonumber\\
&=&2\nabla_d\psi^{bdca}\nabla_b\xi_c+2\nabla_d(\psi^{dacb}\nabla_c\xi_b)-2\nabla_d\psi^{dacb}\nabla_c\xi_b-2\nabla_c(\nabla_d\psi^{cadb}\xi_b)-2\nabla_c(\nabla_d\psi^{cadb}\xi_b)+2\nabla_d\psi^{cadb}\nabla_c\xi_b+2\nu_\xi^a\nonumber\\
&=&2\nabla_c(\psi^{cadb}\nabla_d\xi_b-2\nabla_d\psi^{cadb}\xi_b)+2\nu_\xi^a,
\end{eqnarray}
where $\psi^{[abc]d}=0$ is used in the second and fourth steps, and $\nu_\xi^a=\xi_b\mathcal{I}^{ab}$ with
\begin{equation}
\mathcal{I}^{ab}=\frac{\partial F}{\partial g_{ab}}+2P^{aefd}\varrho^b{}_{efd}.
    \end{equation}
This implies that 
\begin{equation}
\mathbf{J}_\xi-\bm{\mathcal{C}}_\xi=d\mathbf{Q}_\xi +2\mathbf\nu_\xi\cdot\bm\epsilon,
\end{equation}
where 
\begin{equation}
\mathbf{Q}_\xi=(-\psi^{cadb}\nabla_{[d}\xi_{b]}+2\nabla_{[d}\psi^{cadb}\xi_{b]})\bm{\epsilon}_{ca\cdot\cdot\cdot}.
\end{equation}
Note that $d(\mathbf \nu_\xi\cdot\bm\epsilon)=0$ for an arbitrary vector field $\xi$ further gives rise to $\mathcal{I}^{ab}=0$\footnote{We thus provide a simple derivation of this magic identity, which is alternative to that presented in \cite{pa}.}, so we end up with 
\begin{equation}
    \mathbf{J}_\xi-\bm{\mathcal{C}}_\xi=d\mathbf{Q}_\xi.
\end{equation}

\section{The Christoffel symbol in the Gaussian null coordinates and leading behavior of $\nabla_a n_b$}\label{leading}
Here we would like to write down the explicit expression of the Christoffel symbol in the Gaussian null coordinates for the neighbourhood of the event horizon of the stationary black holes as follows
\begin{eqnarray}
\Gamma^{u}\,_{u u} &=& \Gamma^{v}\,_{u u} =\Gamma^{v}\,_{u v} =\Gamma^{v}\,_{vu} =\Gamma^{v}\,_{u i} = \Gamma^{v}\,_{iu}=\Gamma^{i}\,_{u u}=0,\nonumber\\
\Gamma^{u}\,_{u v} &=& \Gamma^{u}\,_{vu} =-\frac{u}{2} [2\alpha  +w^2 +r \partial_{r}(\alpha +\frac{1}{2}{w}^{2})],\nonumber\\
\Gamma^{u}\,_{u i} &=&\Gamma^{u}\,_{iu}= \frac{1}{2}w_{i} +\frac{r}{2} (\partial_{r}{w_{i}}  - w^{j} \partial_{r}{\gamma_{i j}}),\nonumber\\
\Gamma^{u}\,_{v v} &=& {u}^{3} [ -w^{i} \partial_{r}{w_{i}} +{w}^{2} \alpha +{\alpha}^{2}  - \frac{1}{2}w^{i} \partial_{i}{\alpha} +\frac{1}{2} (r({w}^{2} +\alpha )-1)\partial_{r}{\alpha}],\nonumber\\
\Gamma^{u}\,_{v i} &=&\Gamma^{u}\,_{iv}=- \frac{u^2}{2} [ \partial_{i}{\alpha}  +({w}^{2}   +\alpha) (w_{i}+r\partial_r w_i) +w^{j} ( 2\partial_{[i}{w_{j]}}  +\partial_{r}{\gamma_{i j}} ) ],\nonumber\\
\Gamma^{u}\,_{i j} &=& \frac{u}{2} [2\partial_{(i}{w_{j)}}  - \partial_{r}{\gamma_{i j}} +w^{k} ( -2 \partial_{(i}{\gamma_{j) k}} +\partial_{k}{\gamma_{i j}}) -r (  {w}^{2}  + \alpha )\partial_{r}{\gamma_{i j}}],\nonumber\\
\Gamma^{v}\,_{v v} &=& u (\frac{1}{2}r \partial_{r}{\alpha} +\alpha),\nonumber\\
\Gamma^{v}\,_{v i} &=& \Gamma^{v}\,_{iv}=-\frac{1}{2}(w_i+f\partial_f w_i),\nonumber\\
\Gamma^{v}\,_{i j} &=&  - \frac{1}{2}v \partial_{r}{\gamma_{i j}},\nonumber\\
\Gamma^{i}\,_{u v}& =&\Gamma^{i}\,_{vu}= \frac{1}{2}(w^{i} +r \gamma^{i j} \partial_{r}{w_{j}}),\nonumber\\
\Gamma^{i}\,_{v v} &=& {u}^{2} (\gamma^{i j} \partial_{r}{w_{j}} -\alpha w^{i} +\frac{1}{2}\gamma^{i j} \partial_{j}{\alpha}  - \frac{1}{2}r \partial_{r}{\alpha} w^{i}),\nonumber\\
\Gamma^{i}\,_{u j} &=& \Gamma^{i}\,_{ju} =\frac{1}{2}v\gamma^{i k} \partial_{r}{\gamma_{j k}},\nonumber\\
\Gamma^{i}\,_{j k} &=& \frac{1}{2}(\gamma^{i l} \partial_{k}{\gamma_{j l}} +\gamma^{i l} \partial_{j}{\gamma_{k l}}  - \gamma^{i l} \partial_{l}{\gamma_{j k}} +r w^{i} \partial_{r}{\gamma_{j k}}),
\end{eqnarray}
whereby the leading behavior for $\nabla_a n_b$ with $r\rightarrow 0$ along the $t=0$ surface can be obtained as follows
\begin{eqnarray}
\nabla_{u} n_{u}& =& - {\lambda}^{3} v^2[\frac{1}{2}{r}^{2} \partial_{r}(w^2+\alpha)+r (w^2+\alpha)+1]+\lambda v  \Gamma^{u}\,_{u u}+ \lambda u\Gamma^{v}\,_{u u}\xrightarrow [r\rightarrow 0]{t=0}-\frac{1}{2 }\lambda+\mathcal{O}(r^{1/2}) ,\nonumber\\
\nabla_{u} n_{v}& =& \nabla_{v} n_{u}= \lambda-{\lambda}^{3} r[   \frac{1}{2}{r}^{2}\partial_{r}(w^2+\alpha)+r (w^2+\alpha)+1]-\lambda v \Gamma^{u}\,_{u v}-\lambda u \Gamma^{v}\,_{u v}\xrightarrow[r\rightarrow 0]{t=0}\frac{1}{2}\lambda+\mathcal{O}(r^{1/2}) ,\nonumber\\
 \nabla_{u} n_{i}&=& -\lambda v \Gamma^{u}\,_{u i} -\lambda u \Gamma^{v}\,_{u i}\xrightarrow[r\rightarrow 0]{t=0}\mathcal{O}(1),\nonumber\\ 
  \nabla_{i} n_{u}&=&-\frac{1}{2}{\lambda}^{3} {r}^{2} v \partial_{i}(w^2+\alpha) -\lambda v \Gamma^{u}\,_{u i} -\lambda u \Gamma^{v}\,_{u i}\xrightarrow[r\rightarrow 0]{t=0}\mathcal{O}(1),\nonumber\\
 \nabla_{v} n_{v} &=& -\frac{1}{2}{\lambda}^{3}u^2 [{r}^{2}  \partial_{r}(w^2+\alpha)+r (w^2+\alpha)+1]-\lambda v \Gamma^{u}\,_{v v}-\lambda {u} \Gamma^{v}\,_{v v}\xrightarrow[r\rightarrow 0]{t=0}-\frac{1}{2}\lambda+\mathcal{O}(r^{1/2}),\nonumber\\
 \nabla_{v} n_{i}& = & - \lambda v\Gamma^{u}\,_{vi} -\lambda u \Gamma^{v}\,_{vi}\xrightarrow[r\rightarrow 0]{t=0}\mathcal{O}(1) ,\nonumber\\
  \nabla_{i} n_{v}& = & -\frac{1}{2}{\lambda}^{3} {r}^{2} u  \partial_{i}(w^2+\alpha )- \lambda v\Gamma^{u}\,_{vi} -\lambda u \Gamma^{v}\,_{vi}\xrightarrow[r\rightarrow 0]{t=0}\mathcal{O}(1) ,\nonumber\\
 \nabla_{i} n_{ j} &=& -\lambda v \Gamma^{u}\,_{i j} -\lambda u \Gamma^{v}\,_{i j}\xrightarrow[r\rightarrow 0]{t=0}\mathcal{O}(r^{1/2}).
\end{eqnarray}

\end{document}